\documentclass[aps,prd,final,twocolumn,floatfix]{revtex4}
\usepackage{graphicx}
\def\be{\begin{equation}}
\def\ee{\end{equation}}
\def\ba{\begin{array}}
\def\ea{\end{array}}
\def\bea{\begin{eqnarray}}
\def\eea{\end{eqnarray}}

\def\({\left(}
\def\){\right)}
\def\[{\left[}
\def\]{\right]}
\begin{document}
\title{Time dependent non-Abelian waves and their stochastic regimes for gauge fields coupled to external sources}
\author{T. Shreecharan}
\email{shreecharan@ifheindia.org}
\address{Department of Physics, Faculty of Science and Technology, ICFAI Foundation for Higher Education \\ 
Dontanapally, Hyderabad 501203, Telangana, India.}
\author{Thokala Soloman Raju}
\email{tsraju@ifheindia.org}
\address{Department of Physics, Faculty of Science and Technology, ICFAI Foundation for Higher Education \\ 
Dontanapally, Hyderabad 501203, Telangana, India.}
\begin{abstract}
In this paper we explore explicit exact solutions of the $SU(2)$ Yang-Mills (YM) and Yang-Mill-Higgs (YMH) equations with homogeneous and inhomogeneous external sources. Whereas in the case of YM we have confirmed our analytical findings with the numerical simulations, the numerical corroborations in the YMH case yielded the stochastic character of motion for the ensuing fields.
\end{abstract}
\maketitle
\section{Introduction}

It is well-known that the non-Abelian YM gauge theory is essentially nonlinear in nature. More specifically the infrared
phenomena of QCD such as confinement and chiral symmetry
breaking can be explained, using nonperturbative methods, due to the nonlinear nature of gluon interactions. The nonperturbative methods that have been developed over the years have greatly benefited from the study of exact solutions of the classical YM equations \cite{actor}. Of the several exact solutions that have been found for these dyanmical equations of YM and YMH, the most prominent ones are instantons \cite{atiyah,brihaye,shifman}, merons \cite{furlan1,furlan2}, vortices \cite{nielsen}, monopoles \cite{hooft,polyakov,prasad}, and non-Abelian plane waves \cite{coleman,lo1,lo2,savvidi1,savvidi2,savvidi3,savvidi4}.

The pursuit of finding exact solutions for these dynamical equations becomes much more arduous in the presence of external sources. In the literature limited works have been devoted to this task, despite the fact that addition of external sources enriches the dynamics of the ensuing fields. However even with the relative dearth of exact solutions of YM with external sources, some interesting results have been reported \cite{weiss,oh1,braghi}.
Furthermore, the investigation of general properties of classical
solutions of gauge field equations is important and may provide
new insight about the vacuum structure of a given
theory. Out of these properties, the study of irregular stochastic character
of the gauge field dynamics is of great interest. In their seminal papers \cite{savvidi1,savvidi2,savvidi3,savvidi4}, the authors have studied the stochastic behavior of the YM theory. Also the chaotic
behavior of the YM mechanics with 3 degrees of freedom
was demonstrated by a Painleve´ test \cite{steeb1} and by studying
Lyapunov exponents \cite{steeb2}. Additionally, it will be very interesting to consider extra possible interactions
with a Higgs field. In Refs. \cite{savvidi2,kumar,kawabe1,kawabe2} investigations of the
stochastic behavior of the Abelian and non-Abelian classical field systems with Higgs fields but without a source term were made. 

Motivated by the above works, we have discovered exact solutions for the YM and YMH dynamical equations in the presence of external sources. Two special cases have been considered: In the first case, we have explored exact solutions of YM field equations when the external sources are homogeneous. The exact solutions are explicated in detail for different parameter values. In the case of YM equations of motion, numerical simulations conducted attest our analytical results. We have also found exact soliton solutions when the source is inhomogeneous. The phase space portraits have been obtained when the source is an oscillatory one.  In the second case, we have obtained nondegenerate soliton solutions of the YMH dynamical equations, when the source is homogeneous. Parameter domains are delineated in which the solitons exist. Also we have obtained phase space portraits when the sources are nondegenrately inhomogeneous, indicative of stochastic motion.

\section{The Yang-Mills equations of motion}

Here we consider $4-D$ $\ SU(2)$ YM Lagrangian density coupled to an external source, in Minkowski space with the metric $g_{\mu \nu} = \mathrm{diag} (+,-,-,-)$. The Lagrangian in this case is given by
\begin{equation}
\mathcal{L} = - \frac{1}{4} F_{\mu \nu}^a F^{\mu \nu a} + A_\mu^a J^{\mu a}
\end{equation} 
where 
\begin{equation}
F_{\mu \nu}^a = \partial_\mu A_\nu^a - \partial_\nu A_\mu^a + g \epsilon^{abc} A_\mu^b A_\nu^c
\end{equation}
with $\epsilon^{abc}$ being the $SU(2)$ structure constants.  The equations of motion are
\begin{equation} \label{eomym}
\partial_\mu F^{\mu \nu a} + g \epsilon^{abc} A_{\mu}^b F^{\mu \nu c} = J^{\nu a}
\end{equation}
As discussed in Ref. \cite{savvidi1} it is convenient to solve these equations (\ref{eomym}) in the gauge: $A_0^a$, $\partial^i A_i^a=0$. In the next section we would like to explicate the procedure to obtain time dependent solutions.

 We seek solutions that are purely time dependent. To achieve this the ansatz \cite{ebert} is chosen to be :
\begin{equation}
A_{\mu}^a = \delta_{\mu}^a f_{a}(t), \quad A^{\mu a} = g^{\mu a}f_{a}(t), \quad J^{\mu a} = \delta^{\mu a}j_{a}(t),
\end{equation}
where $ a=1,2,3, \, \mu =1,2,3,4\ (x_1=t , x_2 = x , x_3 = y, x_4=z)$. The equations of motion to Eq. (\ref{eomym}) can be written in explicit component form as
\begin{eqnarray}
g^2 f_1(f_2^2 +f_3^2) = j_1 , \nonumber \\
-\ddot{f_2} + g^2 f_2(f_1^2 -f_3^2) = j_2 , \nonumber \\
-\ddot{f_3} + g^2 f_3(f_1^2 -f_2^2) = j_3 ,\\
g(\dot{f_1}f_3 + 2f_1\dot{f_3}) = 0 , \nonumber \\
g( \dot{f_1}f_2 + 2f_1\dot{f_2}) = 0 , \nonumber
\end{eqnarray}
where $\dot{f_{a}}$ denotes derivative with respect to time. One can notice, that this system has only finite number of degrees of freedom, and hence finite dimensional phase space. Choosing $f_{1}=0$ implies $j_1=0$, and further setting $f_2 = f_3 = f(t)$,  $j_2=j_3=j(t)$ in the above set of equations, we obtain
\begin{equation} \label{elliptic1}
f^{\prime \prime} + g^2 f^3 + j(t) = 0 \ .
\end{equation}

It is interesting to note that Eq. (6) without source was derived in Ref. \cite{treat}. 
In this case the energy integral of motion can be written as
\begin{equation}\label{eint1}
f^{\prime \ 2} + \frac{g^2}{2} f^4 + 2 j f = \mathcal{E} \ .
\end{equation}
The above equation describes an anharmonic oscillator with potential $U = \frac{g^2}{2} f^4 + 2 j f$. Although Eq. (\ref{eint1}) can be integrated to yield elliptic functions as solutions in the absence of the source, previously we have shown that there exists M\"{o}bius transform solutions to Eq. (\ref{elliptic1}) when the source is homogeneous in nature i.e., $j(t) = j$ \cite{vivek}.

\begin{figure}
\centering
\includegraphics[width=3in,height=2in]{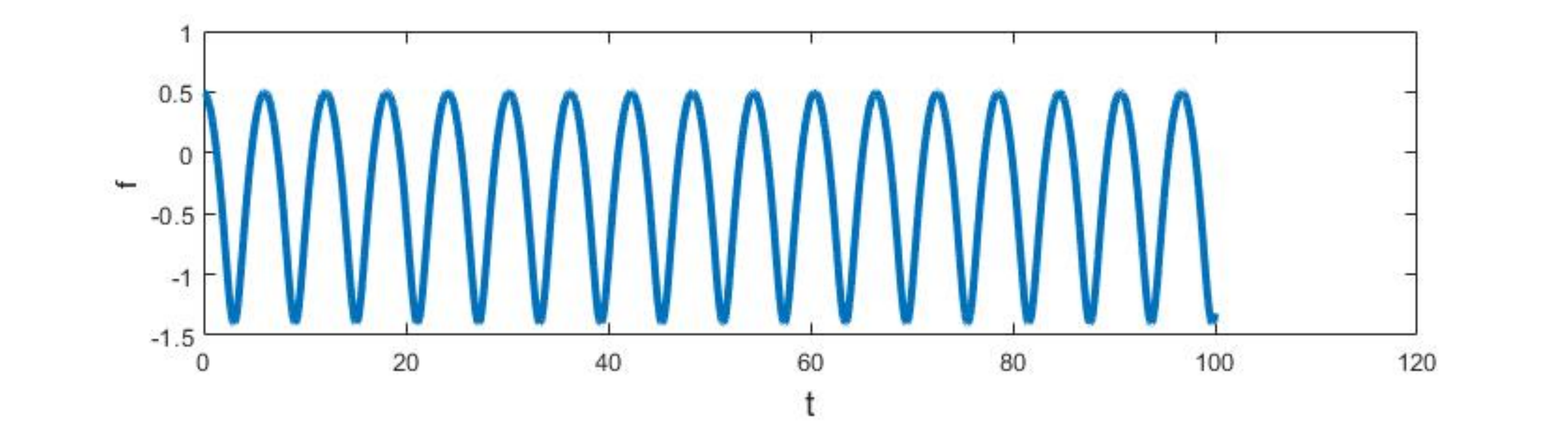}
\caption{Numerical solution for initial conditions $f_{0}=0$, $\dot{f}_{0}=0.5$, $g=1$, and $j=0.5.$}\label{numsol1}
\end{figure}

\begin{figure}
\centering
\includegraphics[width=3in,height=2in]{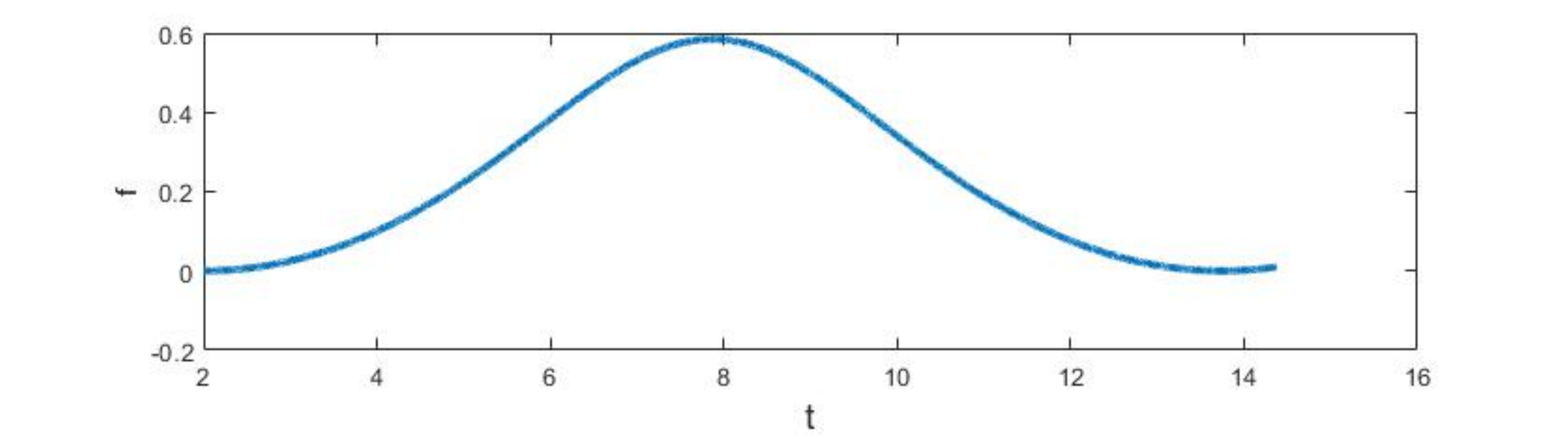}
\caption{Numerical solution for initial conditions $f_{0}=2 \ \dot{f}_{0}=0$, $g=1$, and $j=1.5.$}\label{numsol2}
\end{figure}

Here we will numerically analyze the behavior of the effective mechanical system with the potential $U(f,j)$ leading to Eq. (\ref{elliptic1}). We have solved Eq. (\ref{elliptic1}) numerically using RK-4 technique for two different sets of initial conditions. From the Figs. (\ref{numsol1}) and (\ref{numsol2}), one can see that there exits periodic solutions and soliton solutions for different initial conditions specified in the figure captions. In Fig. (\ref{numsol1}) we depict the numerically obtained periodic solution when the strength of the source is $j=0.5$ and in Fig. (\ref{numsol2}), we depict soliton solution when the strength of the source is $j=1.5$.
In addition to the above solutions we have also found time dependent non-Abelian waves in terms of Jacobian elliptic functions when the external source is inhomogeneous. These are nonperturbative exact solutions.
\newline
\noindent \textbf{Case I:} When the source is $j(t) = B \ \mathrm{cn}(\alpha t,k)$ we find that the solution to Eq. (\ref{elliptic1}) takes the form $f= A \ \mathrm{cn}(\alpha t, k)$, where the amplitude parameters are $A = \sqrt{2 \alpha^2 k/g^2}$, $B = \alpha^2 (1-2k) \sqrt{2 \alpha^2 k/g^2}$.

\begin{figure}
\centering
\includegraphics[width=3in,height=2in]{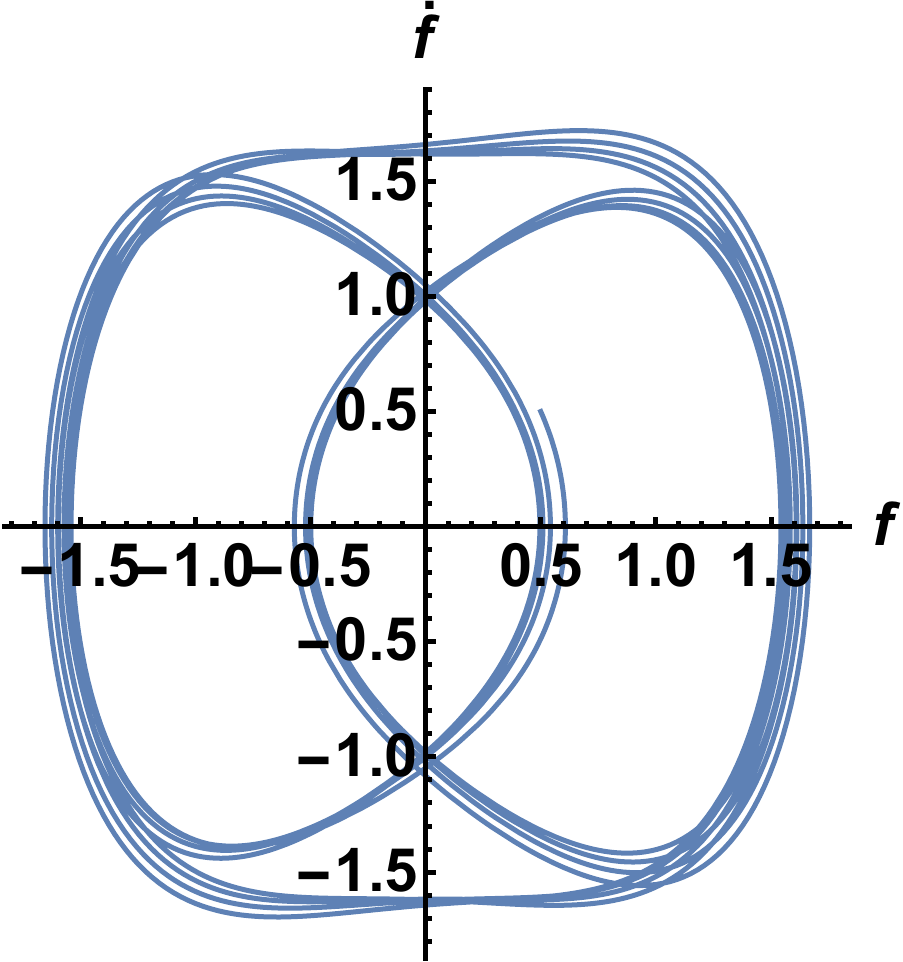}
\caption{Phase portrait for the YM field with an external source $j(t) = \mathrm{cn}(0.5t,0.5)$. Initial conditions are $f_0=0.5 \ \dot{f}_0=0.5$ }\label{ym_sourcet1}
\end{figure}
Apart from the above analytical insights we have also solved the Eq. (\ref{elliptic1}) numerically when the source depends on time explicitly. Specifically we have chosen $j(t) = \mathrm{cn}(0.5t,0.5)$. In Fig. (\ref{ym_sourcet1}) we depict the time evolution of the limit cycle for the initial condition specified in the figure caption.
\noindent \textbf{Case II:} Here we find localized explicit solution to Eq. (\ref{elliptic1}) when the source is a bell-type one : $j(t) = B \ \mathrm{sech}(t)$. We find that the solution is $f= A \ \mathrm{sech}(t)$, where the amplitude parameters are $A = \sqrt{2} /g$ and $B = - \sqrt{2}/g$. We observe that this solution is a limiting case of Case I in the limit $k\rightarrow 1,$ which corresponds to a solution with infinite period of the ($cn$) Jacobian elliptic function. Additionally we have also found dark soliton as exact solution to Eq. (\ref{elliptic1}) when the source is $j(t) = B \ \mathrm{tanh}(t)$.

\begin{figure}
\centering
\includegraphics[width=3in,height=2.3in]{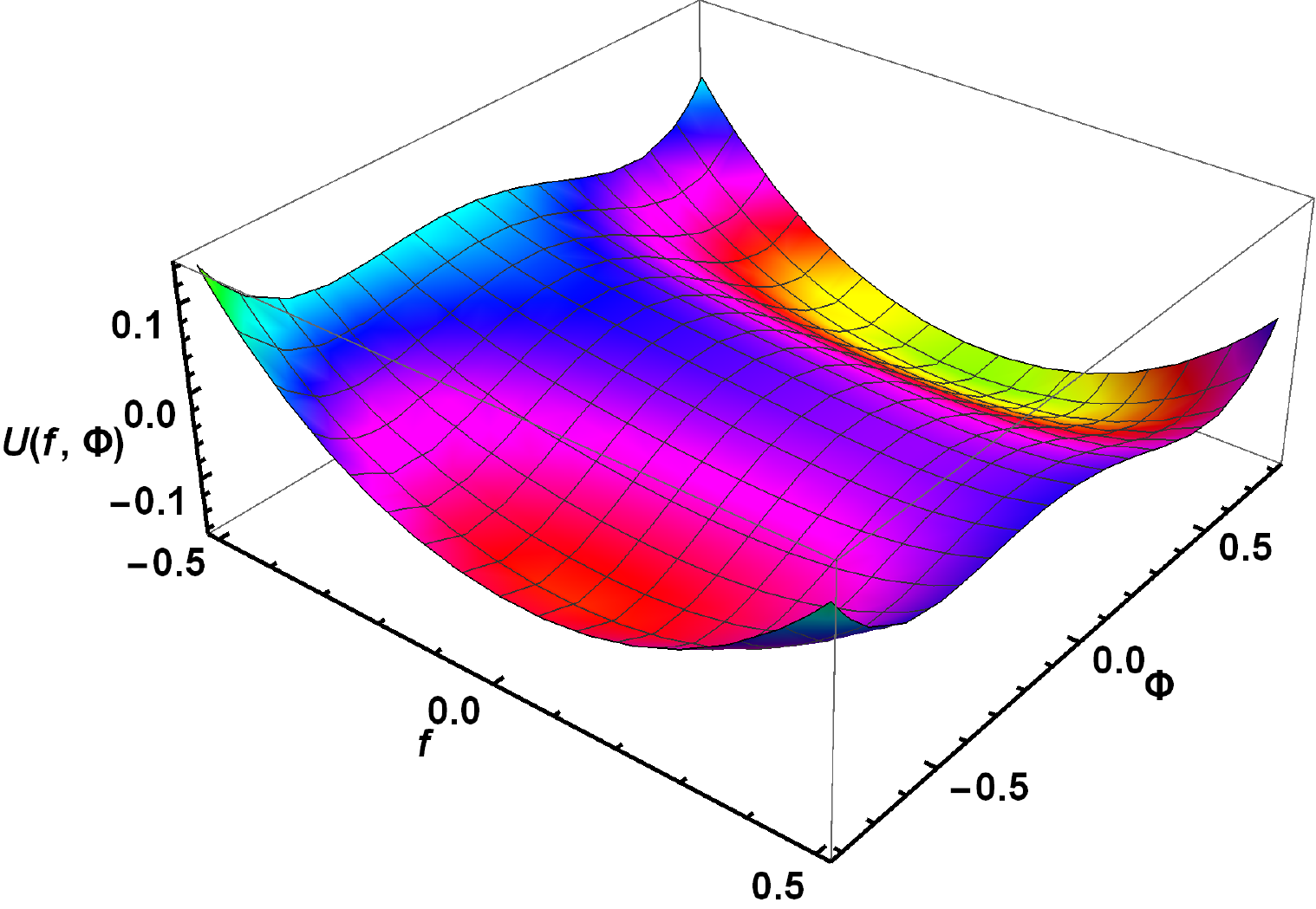}
\caption{Potential $V(f,\Phi)$ for g = 1, $\lambda = 2$, $m = 1$, $j_{1} = 0.05$, and $j_{2} = 0.02$. }\label{pot}
\end{figure}
\begin{figure}
\centering
\includegraphics[width=2.5in,height=1.75in]{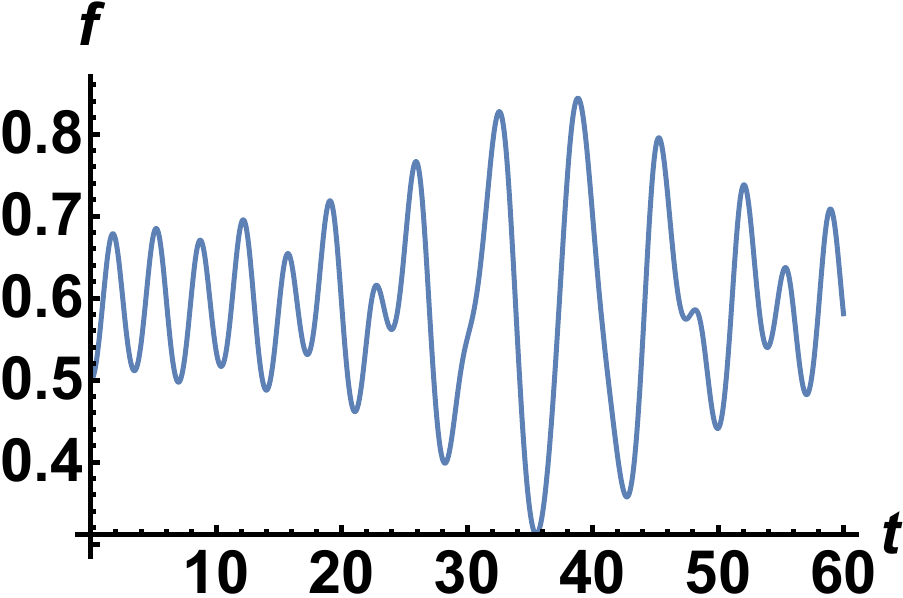}
\caption{System behavior for initial conditions $f_0 = 0.5$, $\dot{f}_0 = 0$, $\Phi_0 = 0.5$, $\dot{\Phi}_0 = 0$ and the set of parameter values $g = 1$, $m=1$, $\lambda = 2$ and $j=0.5$}\label{higgs_source1}
\end{figure}
\begin{figure}
\centering
\includegraphics[width=2.5in,height=1.75in]{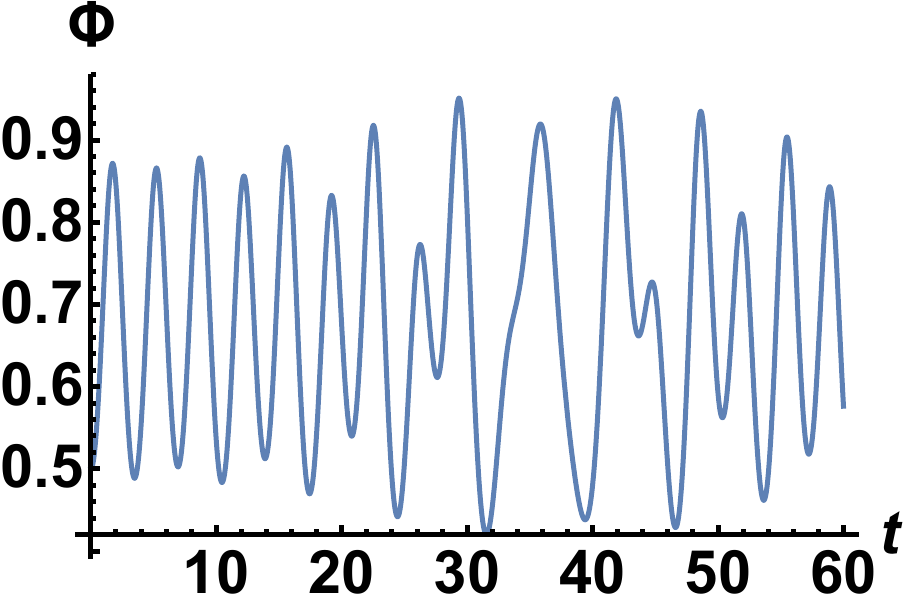}
\caption{System behavior for initial conditions $f_0 = 0.5$, $\dot{f}_0 = 0$, $\Phi_0 = 0.5$, $\dot{\Phi}_0 = 0$ and the set of parameter values $g = 1$, $m=1$, $\lambda = 2$ and $j=0.5$}\label{higgs_source2}
\end{figure}
\section{(3+1)D Yang-Mills-Higgs action}

In the previous section we have looked at the time dependent solutions of the pure YM theory. In this section we generalize the same procedure to the YM field coupled to Higgs field with an external source, specifically we consider the Georgi-Glashow model with an external source.

\begin{figure}
\centering
\includegraphics[width=3in,height=2in]{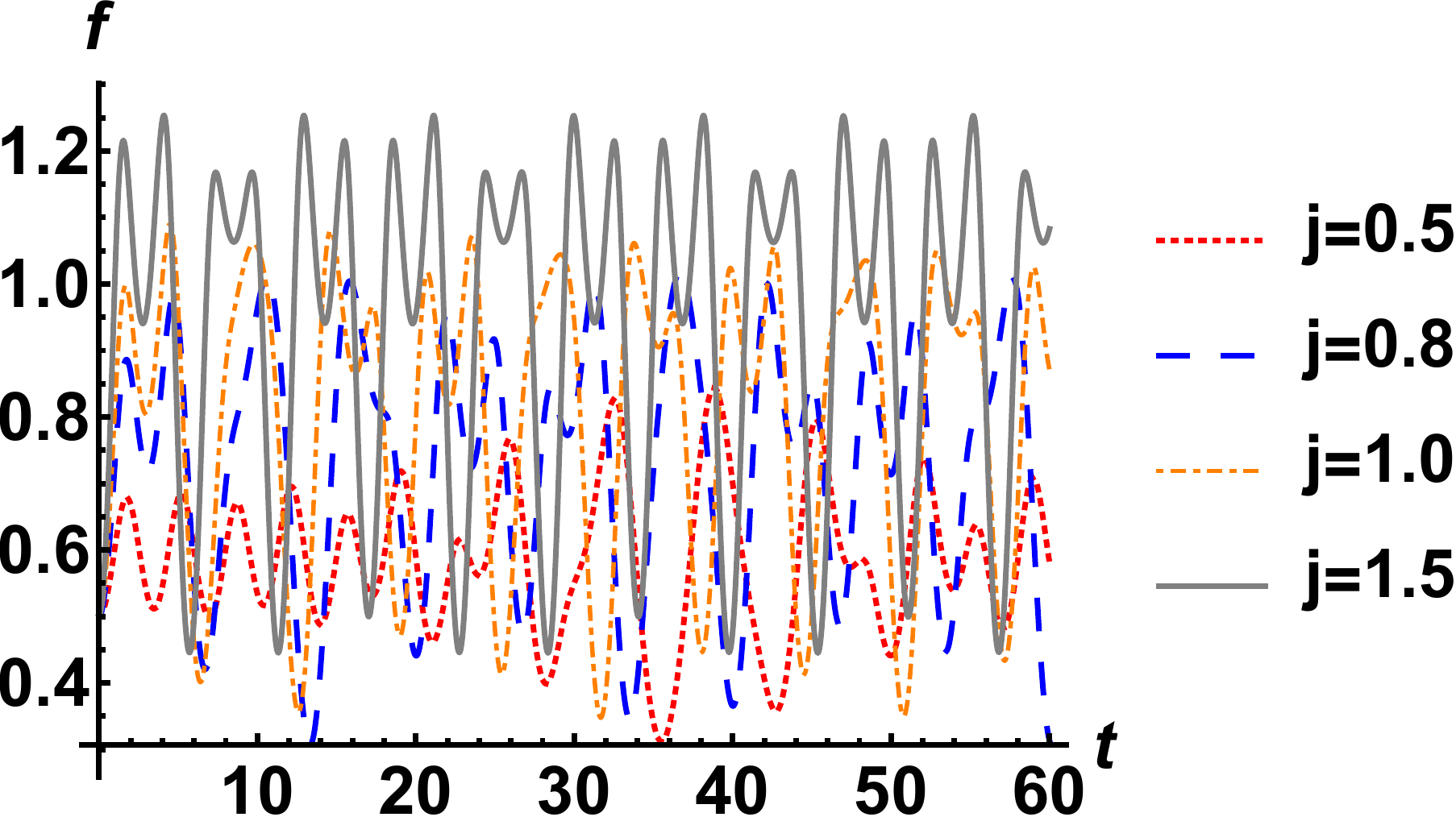}
\caption{System behavior for initial conditions $f_0 = 0.5$, $\dot{f}_0 = 0$, $\Phi_0 = 0.5$, $\dot{\Phi}_0 = 0$ and the set of parameter values $g = 1$, $m=1$, and $\lambda = 2$. Also depicted are the system behavior for various strengths of the sources.}\label{msf}
\end{figure}

\begin{figure}
\centering
\includegraphics[width=3in,height=2in]{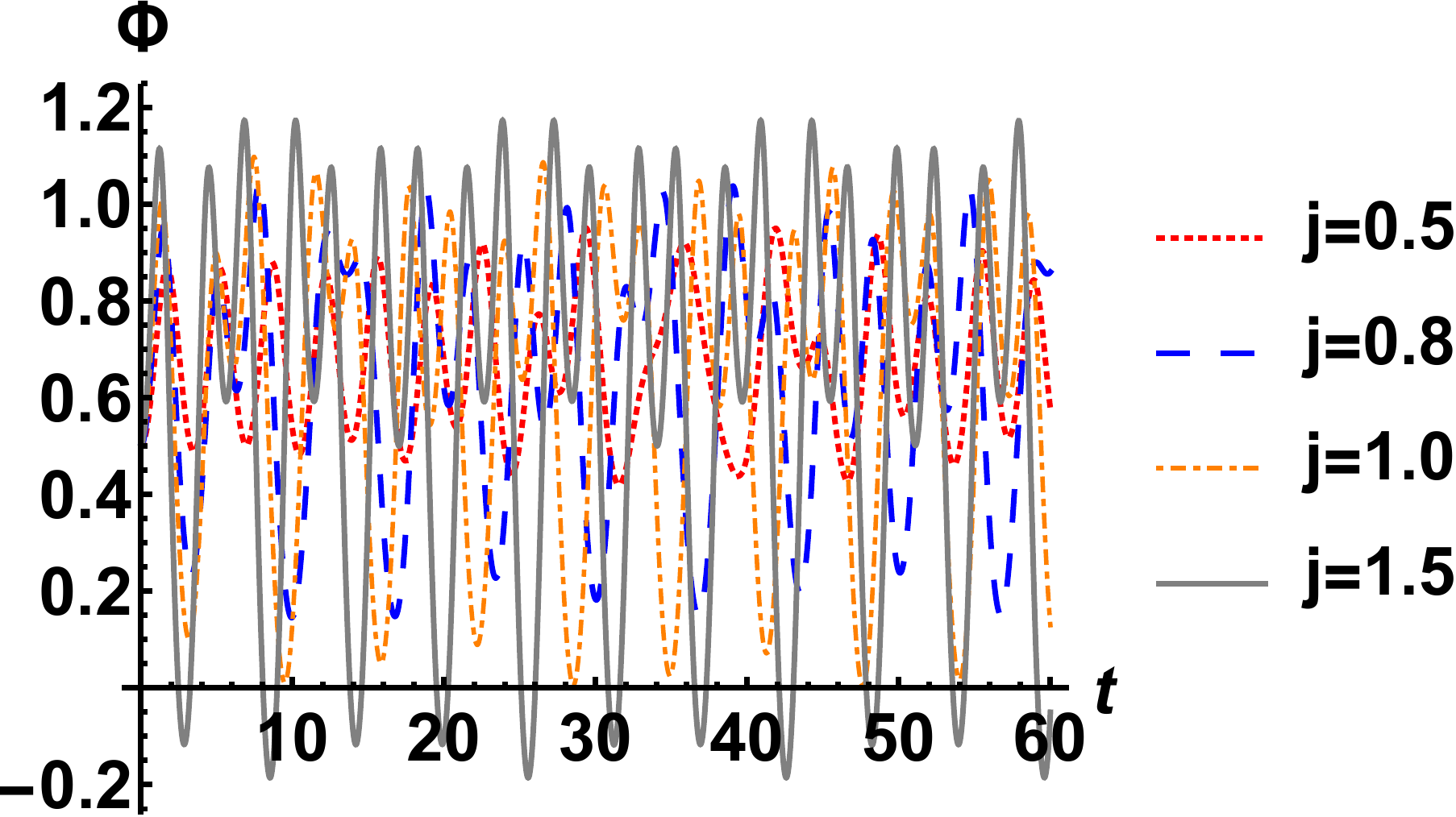}
\caption{System behavior for initial conditions $f_0 = 0.5$, $\dot{f}_0 = 0$, $\Phi_0 = 0.5$, $\dot{\Phi}_0 = 0$ and the set of parameter values $g = 1$, $m=1$, and $\lambda = 2$. Also depicted are the system behavior for various strengths of the sources.}\label{msp}
\end{figure}

\begin{figure}
\centering
\includegraphics[width=2.5in,height=1.75in]{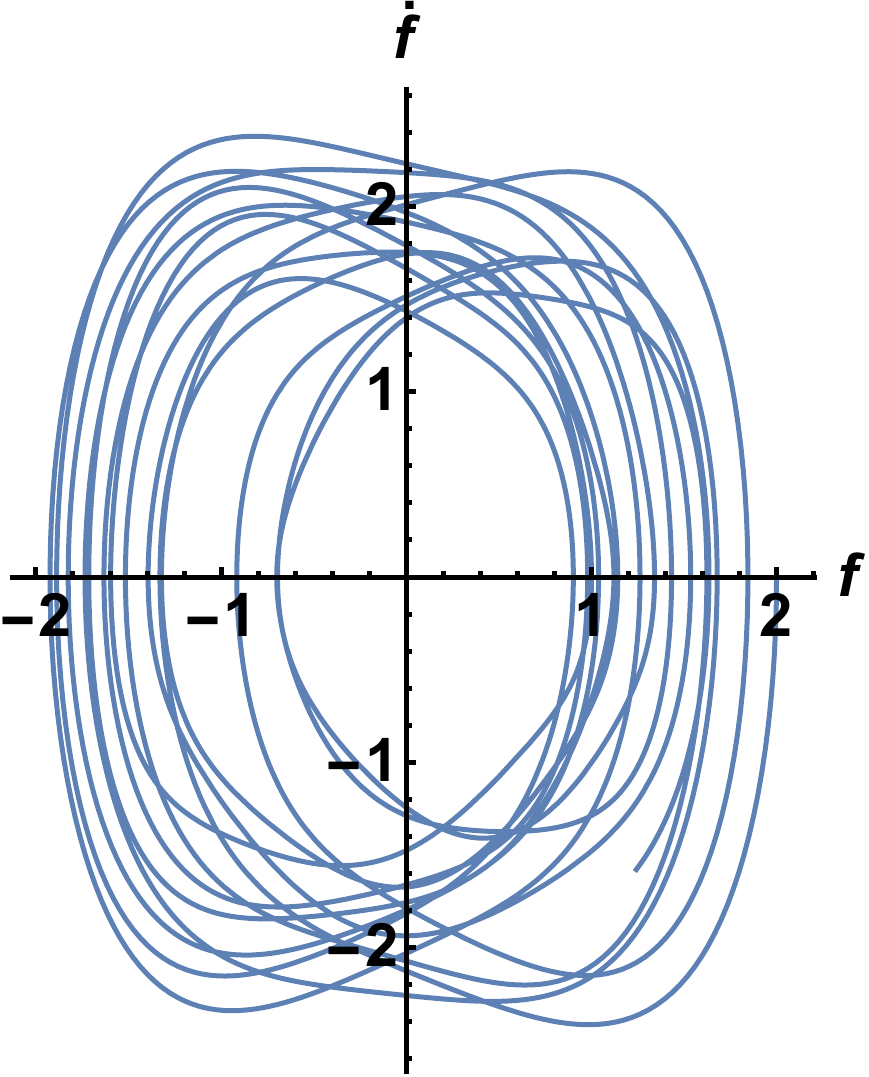}
\caption{Phase space plot for initial conditions $f_0 = 0.5$, $\dot{f}_0 = 0$, $j_{1}(t)=\mathrm{cn}(0.5 t,0.5)$ and the set of parameter values $g = 1$, $m=1$ and $\lambda = 2 $.}\label{higgs_sourcet1}
\end{figure}

\begin{figure}
\centering
\includegraphics[width=2.5in,height=1.75in]{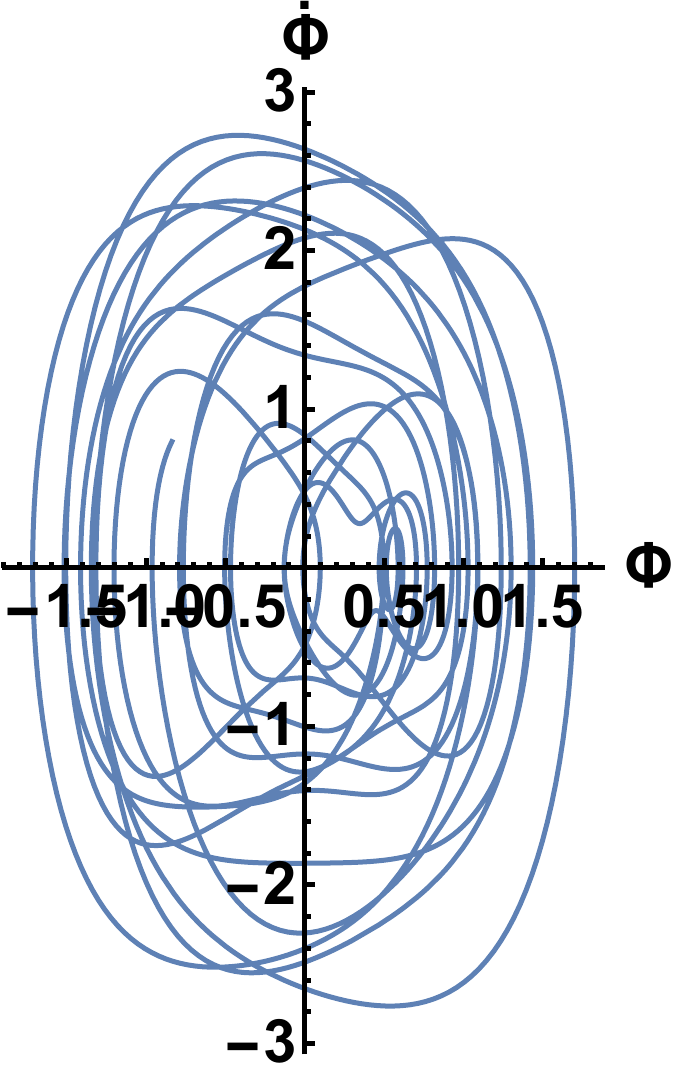}
\caption{Phase space plot for initial conditions $\Phi_0 = 0.5$, $\dot{\Phi}_0 = 0$, $j_{2}(t)=\mathrm{sn}(0.5 t,0.5)$ and the set of parameter values $g = 1$, $m=1$ and $\lambda = 2$.}\label{higgs_sourcet2}
\end{figure}

The Yang-Mills-Higgs (YMH) Lagrangian density coupled to an external source is given by
\begin{eqnarray} \nonumber
\mathcal{L} = - \frac{1}{4} F_{\mu \nu}^a F^{\mu \nu a} + A_\mu^a J^{\mu a} + \frac{1}{2} (D_\mu \Phi^a) (D^\mu \Phi^a) \\ 
+ \frac{m^2}{2} \Phi^a \Phi^a - \frac{\lambda}{4} (\Phi^a \Phi^a)^2 + J^a \Phi^a    
\end{eqnarray}
where the Higgs field is taken to be adjoint thereby $D_\mu \Phi^a = \partial_\mu \Phi^a + g \epsilon^{abc} A_\mu^b \Phi^c$. The equation of motion of the above YMH action is
\begin{eqnarray}
D_\nu F^{\nu \mu a} + g \epsilon^{bac} \Phi^c D^\mu \Phi^b = J^{\mu a} \ ,\\
D_\mu D^\mu \Phi^{a} - m^2 \Phi^a + \lambda (\Phi^b \Phi^b ) \Phi^a = J^{a} \ .
\end{eqnarray}
It is worth mentioning that some algebraic solutions of YMH have been obtained in Ref. \cite{kosinski}.
Here, we follow the procedure enunciated by Ebert et al in Ref \cite{ebert} where the following ansatz has been profitably utilized:
\bea
A_{\mu}^a = \delta_{\mu}^a f_{a}(t), \quad A^{\mu a} = g^{\mu a}f_{a}(t), \quad J^{\mu a} = \delta^{\mu a}j_{a}(t),\nonumber\\
\Phi^{a}(t)=(\Phi_{1}(t),\Phi_{2}(t),\Phi_{3}(t)),
\eea
where $ a=1,2,3, \, \mu =1,2,3,4\ (x_1=t , x_2 = x , x_3 = y, x_4=z)$. In order to seek time dependent solutions, we choose $f_{1}=0$,  $\Phi_{2}=\Phi_{3}=0$ implying $J^{11}=0=J^{23}=J^{32}.$ Further setting $f_2 = f_3 = f(t)$, $\Phi_{1}=\Phi$ we obtain the following coupled equations in gauge and Higgs field:
\begin{eqnarray}
\ddot{f} + g^2 \Phi^2 f + g^2 f^3 = j_1(t) \ , \label{coupled1} \\ \label{coupled2}
\ddot{\Phi} + (2 g^2 f^2 - m^2) \ \Phi + \lambda \Phi^3 = j_2(t) \ ,
\end{eqnarray}
where $J^{22}(t) =J^{33}(t)= j_1(t)$, $J^{1}=0$, and  $J^2(t) =J^3(t)= j_2(t)$. The potential energy of this system is given by
\begin{equation}
V(f,\Phi) = \frac{g^2}{2} f^4 + \frac{\lambda}{4} \Phi^4 - \frac{m^2}{2} \Phi^2 + g^2 f^2 \Phi^2 - j_1 f - j_2 \Phi \ .
\end{equation}
In Fig. (\ref{pot}) we depict this potential energy for different parameters specified in the figure caption. From this figure we observe that there are three critical points given by: $(0,\pm \sqrt{\frac{m+j}{\lambda}})$ and $(0,0)$. It is worth mentioning that there is saddle point with coordinates $(0,0)$.
\begin{figure}
\centering
\includegraphics[width=2.5in,height=1.75in]{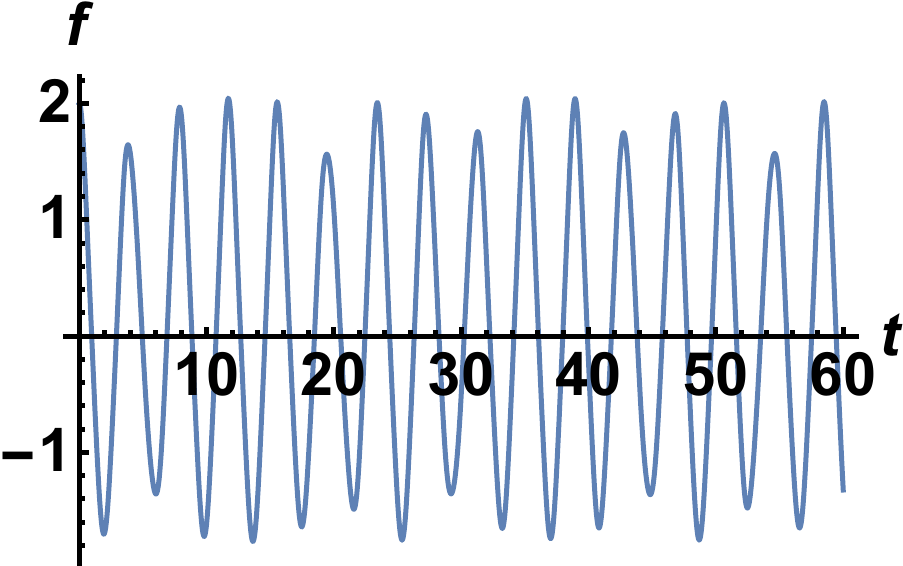}
\caption{System behavior for initial conditions $f_0 = 2.0$, $\dot{f}_0 = 0$, $\Phi_0 = 0.5$, $\dot{\Phi}_0 = 0$ and the set of parameter values $g = 1$, $m=1$, $\lambda = 2$ and $j=0.5$}\label{higgs_source3}
\end{figure}
\begin{figure}
\centering
\includegraphics[width=2.5in,height=1.75in]{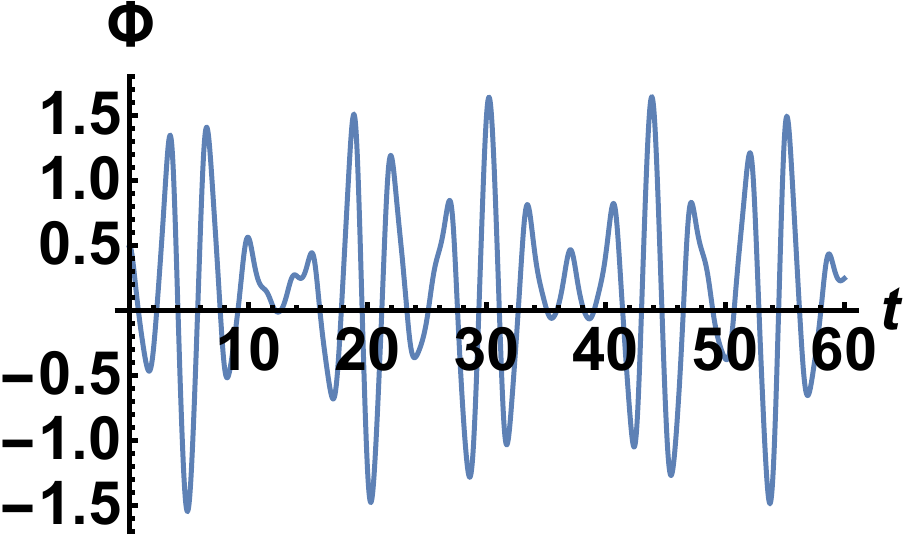}
\caption{System behavior for initial conditions $f_0 = 2.0$, $\dot{f}_0 = 0$, $\Phi_0 = 0.5$, $\dot{\Phi}_0 = 0$ and the set of parameter values $g = 1$, $m=1$, $\lambda = 2$ and $j=0.5$}\label{higgs_source4}
\end{figure}
For the nonlinear coupled equations (\ref{coupled1}) and (\ref{coupled2}), we would like to find nondegenerate non-Abelian waves in terms of Jacobian elliptic functions. We would like to emphasize that these solutions are possible only when the source terms are also Jacobian elliptic functions explicitly depending on time. More pertinently the solutions are:
\begin{eqnarray}
f = A \ \mathrm{cn} (\alpha t, k) \ , \qquad j_1 = B \ \mathrm{cn} (\alpha t, k) \ , \\
\Phi = C \ \mathrm{sn} (\alpha t, k) \ , \qquad j_2 = D \ \mathrm{sn} (\alpha t, k) \ ,
\end{eqnarray}
where the amplitude parameters are given by 
\begin{eqnarray}
A & = & \sqrt{C^2 + \frac{2 k \alpha^2}{g^2}} \ ,\\  
B & = & \Big[g^2 C^2 - \alpha^2 (1-2k) \Big]   \sqrt{C^2 + \frac{2 k \alpha^2}{g^2}} \ , \\
C & = & \sqrt{\frac{2 g^2 A^2 - 2 k \alpha^2}{\lambda}} \ ,\\ 
D & = & \Big[2 g^2 A^2 - \alpha^2 (1+k) - m^2\Big]   \sqrt{\frac{2 g^2 A^2 - 2 k \alpha^2}{\lambda}} . ~~~
\end{eqnarray}
At this moment we will numerically analyze the behavior of the
gauge field and the Higgs field corresponding to the nonlinear coupled equations (\ref{coupled1}) and (\ref{coupled2}). In Figs. (\ref{higgs_source1}) and (\ref{higgs_source2}) we depict the numerical solutions for $f$ and $\Phi$ respectively for one set of initial conditions mentioned in the figure captions. From the figures it is clear that the dynamics of $f$ and $\Phi$ in the presence of a source is much richer as compared to the dynamics without sources Ref. \cite{ebert}. As the strength of the source is increased the motion undergoes a transition from periodic to quasi-periodic and then stochastic. The dynamics for a different set of initial conditions is revealed in Fig. (\ref{higgs_source3}) and (\ref{higgs_source4}). Once again the motion undergoes the transition from periodic to quasi-periodic and then stochastic. These system behaviors have been illustrated in Figs. (\ref{msf}) and (\ref{msp}) by increasing the strength of the external sources for various values mentioned in the figure captions.
\begin{figure}
\centering
\includegraphics[width=2.5in,height=1.75in]{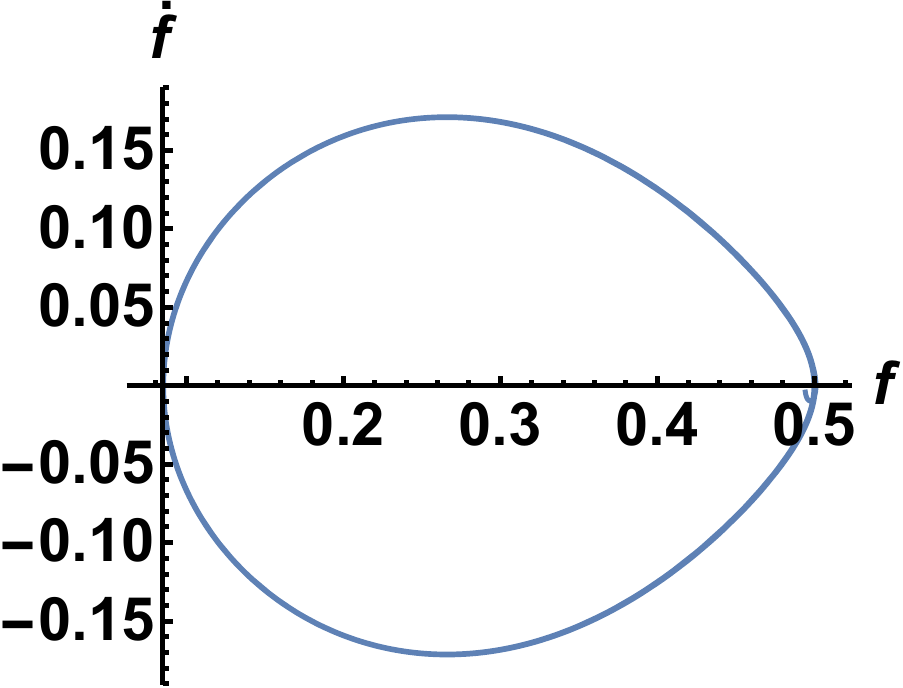}
\caption{Phase space plot for initial conditions $f_0 = 0.5$, $\dot{f}_0 = 0$, $j=0.2$ and the set of parameter values $g = 1$, $m=1$ and $\lambda = 2 $.}\label{paramf1}
\end{figure}
\begin{figure}
\centering
\includegraphics[width=2.5in,height=1.75in]{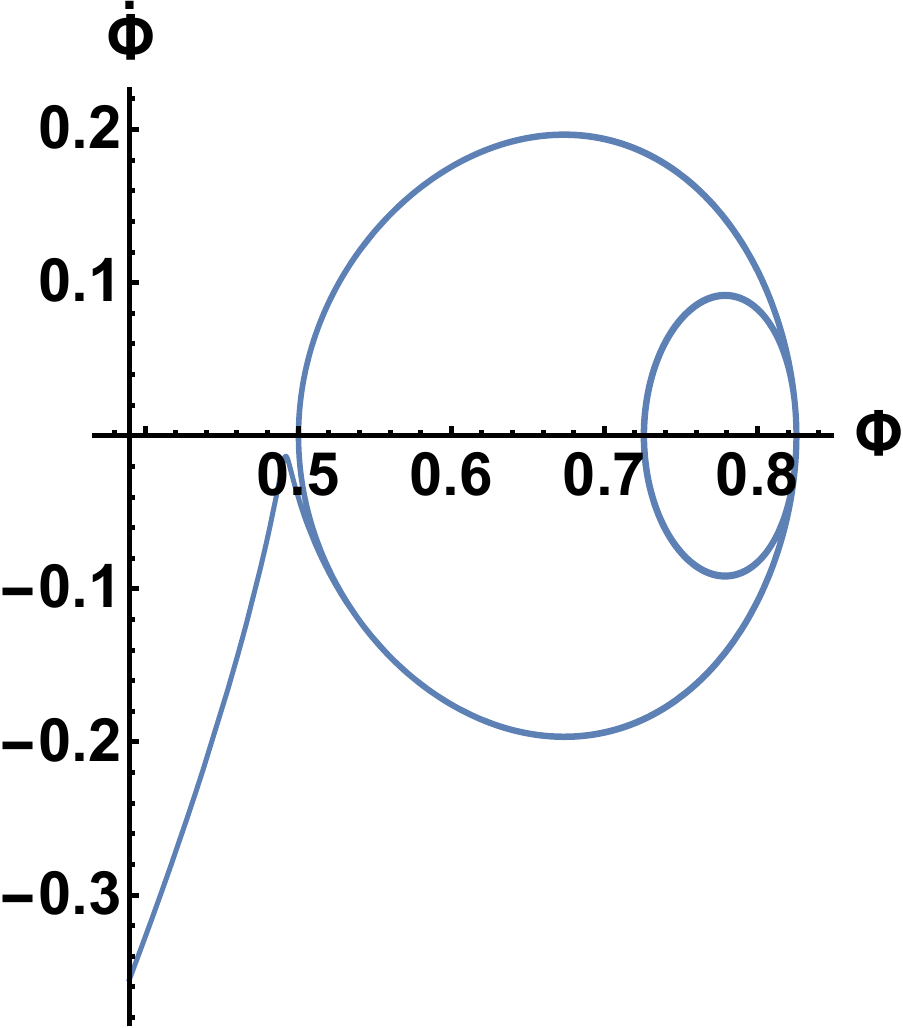}
\caption{Phase space plot for initial conditions $\Phi_0 = 0.5$, $\dot{\Phi}_0 = 0$, $j=0.2$ and the set of parameter values $g = 1$, $m=1$ and $\lambda = 2 $.}\label{paramp1}
\end{figure}
\begin{figure}
\centering
\includegraphics[width=3in,height=2in]{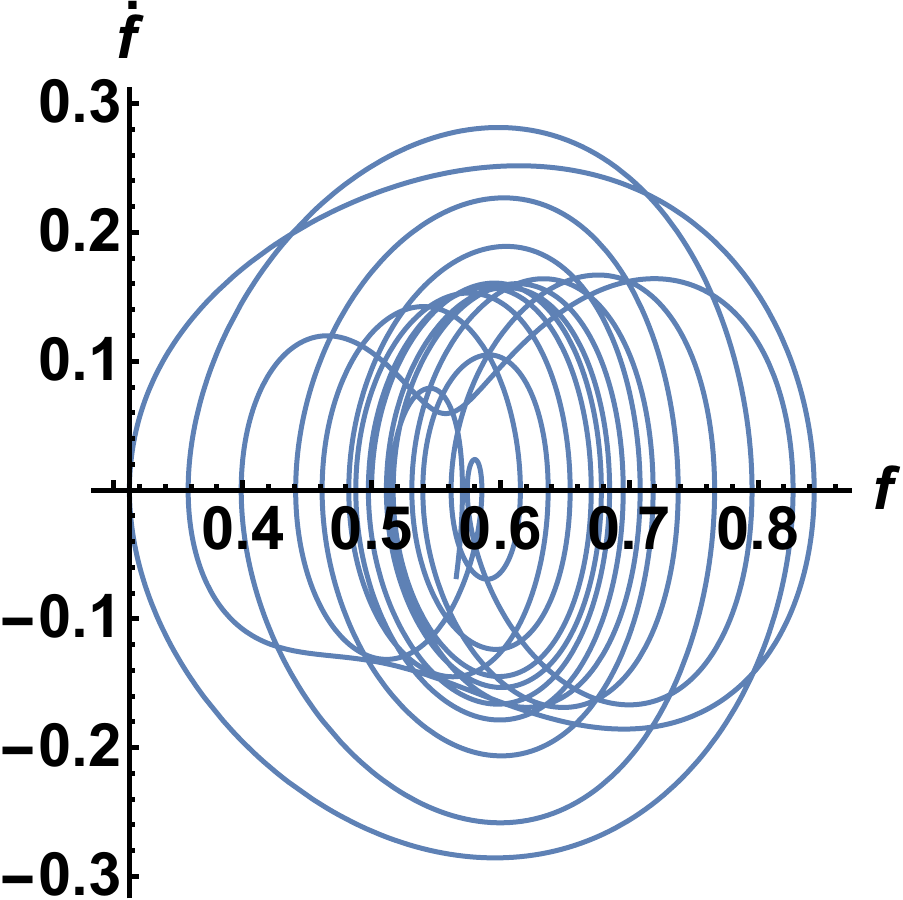}
\caption{Phase space plot for initial conditions $f_0 = 0.5$, $\dot{f}_0 = 0$, $j=0.5$ and the set of parameter values $g = 1$, $m=1$ and $\lambda = 2 $.}\label{paramf2}
\end{figure}
\begin{figure}
\centering
\includegraphics[width=3in,height=2in]{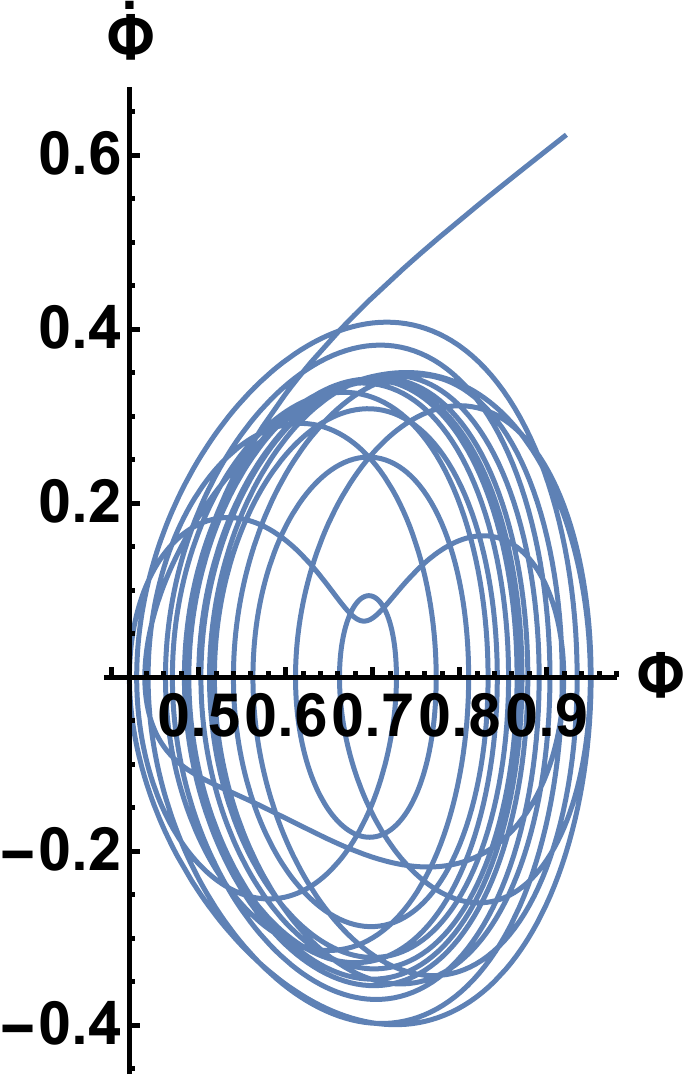}
\caption{Phase space plot for initial conditions $\Phi_0 = 0.5$, $\dot{\Phi}_0 = 0$, $j=0.5$ and the set of parameter values $g = 1$, $m=1$ and $\lambda = 2$.}\label{paramp2}
\end{figure}
In Figs. (\ref{paramf1}) and (\ref{paramp1}) we depict the phase portraits of the chaotic orbit in the $(f-\dot{f})$ and $(\Phi-\dot{\Phi})$ planes for $j=0.2$. As the strength of the source is increased to $j=0.5$, we observe in Figs. (\ref{paramf2}) and (\ref{paramp2}) that the orbit in this case is irregular and densely fills up the region in the phase space. The orbit is evidently nonperiodic and also not quasiperiodic. This orbit is actually chaotic and the region which it occupies is called a chaotic region. In fact, sensitivity with respect to small changes in initial conditions can be verified for the chaotic region of the present system. Before closing this section we would like to append the phase portraits pertaining to the numerical solution of Eqs. (\ref{coupled1}) and (\ref{coupled2}) when the source is non degenerately inhomogeneous. The limit cycles corresponding to the stochastic motion have been depicted in Figs. (\ref{higgs_sourcet1}) and (\ref{higgs_sourcet2}).

\section{conclusion}
In conclusion we have reported exact solutions for the YM and YMH dynamical equations in the presence of external sources. Two special cases have been considered: In the first case, we have explored exact solutions of YM field equations when the external sources are homogeneous. The exact solutions are explicated in detail for different parameter values. In the case of YM equations of motion, numerical simulations corroborated our analytical results. We have also found exact soliton solutions when the source is inhomogeneous. The phase space portraits have been obtained when the source is time dependent Jacobian cosine. In the second case, we have obtained exact nondegenerate soliton solutions of the YMH dynamical equations, when the source is inhomogeneous. We have delineated the parameter domains in which the these solitons exist. Also we have obtained phase space portraits. In this case we have solved the coupled differential equations using RK-4 technique for different initial conditions. The extensive numerical simulations conducted revealed the stochastic motion of the ensuing fields. Despite the fact that the study of the Yang-Mills equations at 
the classical level has already led to numerous remarkable results,  we have still  not yet understood in adequate detail the numerous manifestations of unusual properties of 
the Yang-Mills fields associated with their nonlinear interactions with external sources. Nonetheless, we envision that these presented time-dependent non-Abelian waves will shed new light on our understanding of gluondynamics, involving nonlinear interactions with coupling constant $g$ in the presence of external sources. The external sources that have been used here play a pivotal role in exhibiting the rich dynamical features of both YM and YMH fields. Thus it is hoped that the current work will shed some light in that direction.

\end{document}